\documentclass[pdfusetitle,aps,prd,nofootinbib,twocolumn,superscriptaddress,preprintnumbers]{revtex4-1}
\usepackage{amssymb}
\usepackage{amsmath}
\usepackage{epsfig}
\usepackage{hyperref}
\usepackage{breakurl}
\usepackage{bm}
\usepackage{color}
\usepackage{tikz}
\usepackage[utf8]{inputenc}
\usetikzlibrary{decorations.markings}

\makeatletter
\def\simgt{\mathrel{\lower2.5pt\vbox{\lineskip=0pt\baselineskip=0pt
           \hbox{$>$}\hbox{$\sim$}}}}
\def\simlt{\mathrel{\lower2.5pt\vbox{\lineskip=0pt\baselineskip=0pt
           \hbox{$<$}\hbox{$\sim$}}}}

\def\sectionskip{\vskip .2 cm}

\makeatother

\def\spa#1.#2{\left\langle#1\,#2\right\rangle}
\def\spb#1.#2{\left[#1\,#2\right]}
\def\sand#1.#2.#3{%
\left\langle#1{\vphantom1}\right|{#2}\left|#3\right]}%
\def\sandmp#1.#2.#3{%
\left\langle#1{\vphantom1}\right|{#2}\left|#3\right]}%
\def\sandpm#1.#2.#3{%
\left[#1{\vphantom1}\right|{#2}\left|#3\right\rangle}%
\def\sandmm#1.#2.#3{%
\left\langle#1{\vphantom1}\right|{#2}\left|#3\right\rangle}%
\def\sandpp#1.#2.#3{%
\left[#1{\vphantom1}\right|{#2}\left|#3\right]}%

\renewcommand{\imath}{\mathrm{i}}

\def\Section#1{\noindent {\it #1}}
\newcommand{\be}{\begin{equation}}
\newcommand{\ee}{\end{equation}}
\newcommand{\eq}[2]{\be\begin{aligned}#1 \label{#2}\end{aligned}\ee}

\newcommand{\Eq}[1]{Eq.~\eqref{#1}}

\newcommand{\E}{{\rm E}}
\newcommand{\K}{{\rm K}}

\def\topbotatom#1{\hbox{\hbox to 0pt{$#1\bot$\hss}$#1\top$}} 

\newcommand{\tabeq}[2]{ \parbox{#1}{  \be\begin{aligned}#2 \end{aligned} \nonumber \ee }}

\begin{document}
%\preprint{\preprint{CALT-TH-2021-004, FR-PHENO-2021-03, OUTP-21-03P}}

\title{Scattering Amplitudes, the Tail Effect, and Conservative Binary Dynamics at ${\cal O}(G^4)$}
%\title{The Tail Effect and Conservative Binary Dynamics at ${\cal O}(G^4)$}
%\title{The Post-Minkowskian Tail Effect}
%\title{Complete conservative dynamics at O($G^4$) for binary scattering.}
%want scattering for DOE and our community
%want LIGO as audience
%want mention "tail"

\author{Zvi Bern}
\affiliation{
Mani L. Bhaumik Institute for Theoretical Physics,
University of California at Los Angeles,
Los Angeles, CA 90095, USA}
\author{Julio Parra-Martinez}
\affiliation{Walter Burke Institute for Theoretical Physics,
    California Institute of Technology, Pasadena, CA 91125}
\author{Radu Roiban}
\affiliation{Institute for Gravitation and the Cosmos,
Pennsylvania State University,
University Park, PA 16802, USA}
\author{Michael~S.~Ruf}
\affiliation{
Mani L. Bhaumik Institute for Theoretical Physics,
University of California at Los Angeles,
Los Angeles, CA 90095, USA}
\author{Chia-Hsien Shen}
\affiliation{
Department of Physics, University of California at San Diego, 9500 Gilman Drive, La Jolla, CA 92093-0319, USA}
\author{ Mikhail P. Solon}
\affiliation{
Mani L. Bhaumik Institute for Theoretical Physics,
University of California at Los Angeles,
Los Angeles, CA 90095, USA}
\author{Mao Zeng}
\affiliation{Higgs Centre for Theoretical Physics, University of Edinburgh, James Clerk Maxwell Building, Peter Guthrie Tait Road, Edinburgh, EH9 3FD, United Kingdom}

\begin{abstract}

We complete the calculation of conservative two-body scattering
dynamics at fourth post-Minkowskian order, i.e. ${\cal O}(G^4)$ and
all orders in velocity, including radiative contributions
corresponding to the tail effect in general relativity. As in previous
calculations, we harness powerful tools from the modern scattering
amplitudes program including generalized unitarity, the double copy,
and advanced multiloop integration methods, in combination with
effective field theory. The classical amplitude involves complete elliptic
integrals, and polylogarithms with up to transcendental weight
two. Using the amplitude-action relation, we obtain the radial action
directly from the amplitude, and match the known overlapping terms in
the post-Newtonian expansion.

\end{abstract}
   
\maketitle

\Section{Introduction.}
Gravitational-wave science has opened up a new direction in
theoretical high energy physics: leveraging advances in quantum field
theory (QFT) to develop new tools for the state-of-the-art prediction of
gravitational-wave signals. This era of ever-increasing precision
holds the promise of new and unexpected discoveries in astronomy,
cosmology, and particle physics~\cite{LIGO,NewDetectors}, but hinges
crucially on complementary advances in our theoretical modeling of
binary sources. In recent years, a new program for understanding the
nature of gravitational-wave sources based on tools from scattering
amplitudes and effective field theory (EFT) has emerged.

The central rationale of the new QFT-based approach is to derive
classical binary dynamics from scattering amplitudes in order to take
full advantage of relativity, on-shell
methods~\cite{GeneralizedUnitarity}, double-copy relations between
gauge and gravity theories~\cite{KLT,BCJ}, advanced multiloop
integration~\cite{IBP,DEs,FIRE,PRZ}, and EFT
methods~\cite{Weinberg:1978kz,NeillandRothstein,2PM}. These are the engines that
drive modern scattering amplitude calculations in particle theory, but
had not been fully integrated together and applied for
gravitational-wave signals. This program is rooted in the well-known
connection of scattering amplitudes to general relativity corrections
to Newton's
potential~\cite{ScatteringToGrav,RecentAmpToPotential,NeillandRothstein},
and in the pioneering application of EFT to gravitational-wave
physics~\cite{NRGR}. It also complements traditional
approaches to binary dynamics such as effective one-body~\cite{EOB},
numerical relativity~\cite{NR}, gravitational
self-force~\cite{self_force,SelfForceReview}, and perturbation theory
in the post-Newtonian (PN)~\cite{PN,4PNDJS, 4PNrest, Bini:2017wfr, Blumlein6PN},
post-Minkowskian (PM)~\cite{PM,PMModernList}, and non-relativistic
general relativity~\cite{NRGR,NRGRrecent} frameworks.

Scattering amplitude methods for determining binary dynamics from
potential gravitons have been used in Refs.~\cite{3PM, 3PMLong,
  4PMPotential} to derive state-of-the-art results, which have been
confirmed in multiple studies~\cite{6PNBlumlein, 3PMFeynman,
  3PMworldline, Bini:2020wpo, Bini:2020nsb, Dlapa:2021npj}. Until now,
radiative effects at all orders in velocity have only been considered
at ${\cal O}(G^3)$, where radiation reaction is purely dissipative and
can be cleanly separated and computed~\cite{KMOC,
  UltraRelativisticLimit, HPRZ, OtherVeneziano, HPRZ2,
  BDGcons}. Radiative effects have new features at ${\cal O}(G^4)$,
such as conservative contributions arising from two radiation
gravitons known as the tail effect~\cite{TailEffect,NewTailPapers}, as
well as nonlinear dissipative effects that mix with time-reversal
invariant conservative effects~\cite{BDGcons}.  In this Letter we
extend amplitude methods to include all conservative contributions to
two-body scattering dynamics.

Our complete result for the conservative contributions ultimately
follows from the same condition used to determine the potential
graviton contributions~\cite{3PM, 3PMLong, 4PMPotential}. Thus, while
extracting conservative dynamics in the presence of radiation involves
additional diagrams and expanding integrals in a new region, it is
conceptually no more difficult. Remarkably, at this order in the $G$
expansion, the result obtained with the standard Feynman ${\rm
  i}\varepsilon$ prescription is consistent with using Wheeler-Feynman
time-symmetric graviton propagators, as advocated in
Ref.~\cite{BDGcons}.
 
The classical amplitude presented in Eq.~(\ref{eq:amplitude}) below includes both the radiative contributions found here and the potential contributions from our previous work~\cite{4PMPotential}. It determines the complete conservative
two-body scattering dynamics at ${\cal O}(G^4)$ and all orders in
velocity. Using the amplitude-action relation introduced in
Ref.~\cite{4PMPotential} (see also Refs.~\cite{RadialOthers}), we
determine the radial action in Eq.~(\ref{eq:radial_4}). As usual,
overlapping results from different techniques help guide new
calculations and provide explicit benchmarks. Our result for the ${\cal O}(G^4)$ scattering angle agrees with the sixth PN order result of Ref.~\cite{BDGcons}. 
We also agree with the fifth PN order result of Ref.~\cite{Blumlein:2021txe} up to a single term of higher order in the self-force expansion, whose origin requires further study.

\sectionskip
\Section{Conservative Radiative Effects.} 
The four-point amplitude ${\cal M}$ of gravitationally-interacting
minimally-coupled massive scalars encodes conservative two-body
scattering dynamics in the classical limit.  To compute the amplitude,
we follow textbook QFT rules, such as the use of the Feynman
$\mathrm{i}\varepsilon$-prescription for all propagators. Our task is
then to impose conditions that project out quantum and
dissipative parts, leaving only the contributions to the full classical
conservative dynamics.

We take the classical limit of ${\cal M}$ by expanding in large
angular momentum $J \gg \hbar$~\cite{2PM,3PM,3PMLong}. This amounts to
rescaling the momentum transfer $q$ and all graviton loop momenta
$\ell$ as $q, \ell \to \lambda q, \lambda \ell$ and then expanding in
small $\lambda$.  The classical expansion is thus equivalent to an
expansion in the soft region, defined by the loop-momentum scaling
$\ell = (\omega, \bm \ell) \sim (\lambda, \lambda)$.

%%%%%%%%%%%%%% FIGURE %%%%%%%%%%%%%                             
\begin{figure}[t]
\begin{center}
\includegraphics[scale=.215]{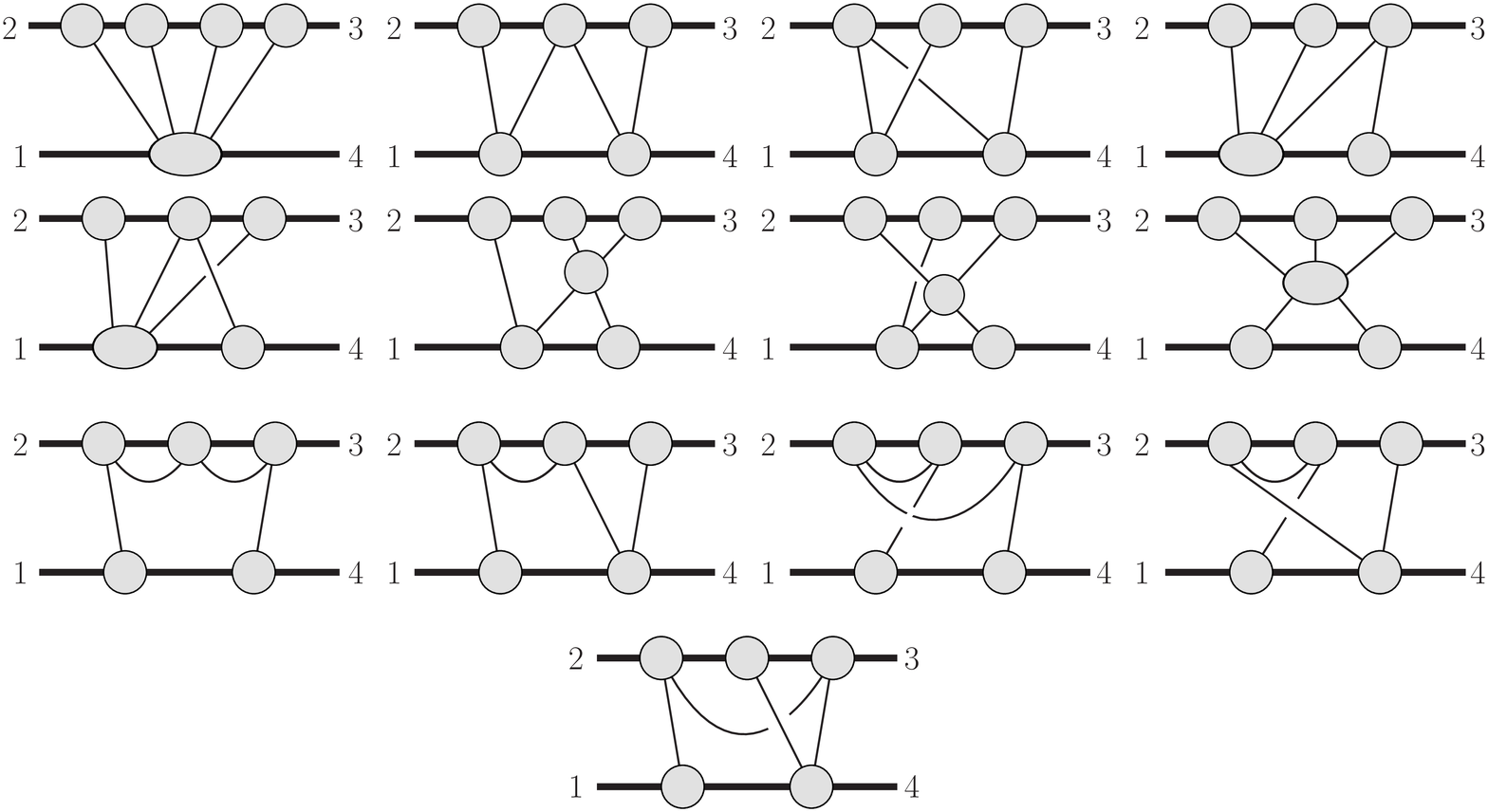}
\end{center}
\vskip -.5cm
\caption{
Complete set of generalized unitarity cuts for conservative contributions at $O(G^4)$.  Exposed
  lines are on-shell. Thick lines represent massive scalars and thin lines are gravitons. The first eight cuts contain both (ppp)
  and (prr) contributions, while the last five, which have gravitons starting and ending on the same matter line, have no (ppp) contributions.
}
\vspace{-0.4cm}
\label{fig:cuts}
\end{figure}
%%%%%%%%%%%%%%%%%%%%%%%%%%%%%%%%%%% 

To define conservative dynamics in the presence of radiative effects,
we evaluate integrals by picking up positive-energy residues of matter
propagators~\cite{2PM,3PM,3PMLong,4PMPotential}.  A key implication of
this is that we only need to consider cuts that have at least one
on-shell matter line per loop.  This is the same condition that
identifies contributions from potential gravitons, and vastly
simplifies the analysis.  The relevant cuts are shown in
Fig.~\ref{fig:cuts}.

Notably, the setup described above assumes the standard Feynman $\mathrm{i}\varepsilon$-prescription for graviton propagators, but is consistent with using a principal-value (PV) prescription corresponding to a time-symmetric propagator~\cite{PVPrescription, DamourPV},
\vskip - .5 cm 
\begin{equation}
G^{\mathrm{PV}}(k)=\mathcal{P}\frac{1}{k^2}\label{eq:PVPrescription}\,,
\end{equation}
where $\mathcal{P}$ denotes the principal value. That is, for the part of the three-loop amplitude
determined by these cuts, the PV prescription gives the same result as using the standard Feynman $\mathrm{i}\varepsilon$-prescription for graviton
propagators and then taking the real part of the final classical amplitude.
Note that the prescription in~(\ref{eq:PVPrescription}) requires some care when applied to multiple propagators~\cite{PVSubtlety}.
%; see Eq.~() of Ref.~\cite{PBtheorem}.

We use the method of regions~\cite{Beneke:1997zp},
separating the soft region into the potential (p) and radiation (r)
subregions defined by the following scalings of the loop momenta
$\ell$:
\begin{equation} 
        \text{(p)}: \ell\sim(v\lambda,\lambda)\,,\quad
\text{(r)}: \ell\sim(v\lambda,v\lambda) \,.
\end{equation} 
Here $v$ denotes the typical velocity of the binary
constituents, corresponding to the small velocity that defines the PN
expansion. 
The classical contribution to any integral is obtained by expanding
the three loop momenta in these regions: $(\ell_1 \ell_2 \ell_3) \sim$
(ppp), (ppr), (prr), and (rrr).  Of these combinations, only (ppp) and (prr)
are relevant for conservative dynamics since (rrr) leads to scaleless integrals and (ppr) yields
odd-in-$v$ contributions, which capture dissipative effects and are
thus removed by the PV-prescription~\eqref{eq:PVPrescription}. Effects
from the square of radiation-reaction, which are dissipative but
even-in-$v$, are also removed by the PV-prescription. The
contributions from the potential region (ppp) have been computed in
our previous work~\cite{4PMPotential} and confirmed in
Refs.~\cite{Blumlein6PN, Dlapa:2021npj}. In this Letter we focus on the remaining
conservative contribution, originating from the (prr) region.

%%%%%%%%%%%%%% FIGURE %%%%%%%%%%%%%                             
\begin{figure}[t]
\begin{center}
\includegraphics[scale=.26]{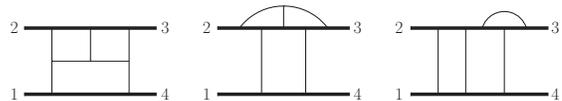}
\end{center}
\vskip -.5cm
\caption{Sample diagrams at ${\cal O}(G^4)$. From left to right: a
  diagram with both (ppp) and (prr) contributions, a diagram with only (prr) contributions, a
  diagram containing iteration terms that cancel in the total amplitude.}
\vspace{-0.4cm}
\label{fig:samplediagrams}
\end{figure}
%%%%%%%%%%%%%%%%%%%%%%%%%%%%%%%%%%% 

\sectionskip
\Section{Constructing the Integrand.} 
As in our previous work~\cite{3PM,3PMLong,4PMPotential}, we use
generalized unitarity and the $D$-dimensional tree-level BCJ double
copy to construct an integrand that captures the complete conservative
dynamics. We sew together tree-level amplitudes in such a way that terms in
the physical state projectors that depend on light-cone reference
momenta drop out automatically~\cite{Kosmopoulos:2020pcd}.  The
relevant radiative contributions come from generalized unitarity cuts
containing one Compton tree amplitude, and two connected five-point
tree amplitudes, each with one cut graviton line to separate the two
matter particles. The subset of these cuts containing gravitons
beginning and ending on the same matter line contribute only to (prr),
while the rest contribute to both (ppp) and
(prr). As before, graphs with graviton bubbles and matter contact
interactions do not contribute. The relevant cuts including both radiative and potential contributions are shown in
Fig.~\ref{fig:cuts}; others are obtained by relabeling the external
momenta.

The resulting integrand is organized in terms of 93 distinct
cubic-vertex Feynman-like diagrams, of which three are shown in
Fig.~\ref{fig:samplediagrams}.
Upon expansion in the soft region, the propagators of certain diagrams, such as the second and third in Fig.~\ref{fig:samplediagrams}, become linearly dependent. This is a feature observed in the
method of regions~\cite{Beneke:1997zp}, and described in
Ref.~\cite{HPRZ2} in the context of
the soft expansion employed here. The products of independent propagators resulting
from partial fraction decomposition can be assigned, upon
multiplication and division by suitable propagators, to 51 distinct
graphs with only cubic vertices.

In contrast to calculations at lower orders, the integrand
at ${\cal O} (G^4)$ depends on whether we use four- or $D$-dimensional
states on cut lines. This is presumably related to subtleties associated with
non-conventional forms of dimensional regularization (see e.g.
Ref.~\cite{DimReductionSubtlety}). To avoid such issues we use
conventional dimensional regularization where all states and momenta
are uniformly continued to $D=4-2 \epsilon$ dimensions.

\sectionskip
\Section{Evaluating the Integrals.}  
The resulting integrals are reduced to a basis of master integrals
using integration-by-parts (IBP) identities~\cite{IBP} and graph symmetries.  An important insight
is that IBP relations are agnostic to the choice of
graviton propagator prescriptions. This allows us to make use of
automated tools such as FIRE6~\cite{FIRE} and KIRA~\cite{KIRA}, which are 
usually applied to cases with Feynman propagators, to reduce integrals
with the prescription \eqref{eq:PVPrescription}.

We evaluate the master integrals using the method of differential
equations~\cite{DEs, PRZ} with boundary conditions determined either
in the (ppp) or (prr) region, and using the PV
prescription~\eqref{eq:PVPrescription}. Since they are disjoint, they
can be treated separately, and the complete system of linear
differential equations splits into two partly-overlapping sectors each
with non-vanishing boundary conditions in one of the two regions. The
overlap is given by integrals with non-trivial boundary conditions in
both regions, which are those captured by the first eight cuts in
Fig.~\ref{fig:cuts}.

The system of differential equations was solved in the (ppp) region in
Ref.~\cite{4PMPotential} in terms of classical polylogarithms up to
weight three and complete elliptic integrals.
The system for the integrals in the (prr) region can be brought to
canonical form~\cite{Henn}, which allows us to evaluate the integrals
to arbitrary order in the dimensional regulator $\epsilon=(4-D)/2$.
The canonical form also implies that elliptic integrals are absent in
the (prr) region.  We have evaluated the integrals up to
transcendental weight three.  Note that due to the requirement of one
on-shell matter line per loop, each of the $L$ loops introduces a
factor of $\pi$, and therefore the maximal weight is $L$ instead of
$2L$ for a general $L$-loop amplitude~\cite{Arkani-Hamed:2010pyv}.
We have also verified that our integration method is consistent with the
nonrelativistic integration approach~\cite{2PM,3PM,3PMLong}.

\sectionskip
\Section{Amplitude.}  The final result for the ${\cal O}(G^4)$ conservative scattering amplitude in the classical limit including all contributions is
\vskip -.7 cm 
\begin{widetext}
	\begin{align}
	{\cal M}_4^{\rm cons} &= 
	G^4 M^7 \nu^2 |\bm q|\pi^2 \left[ {\cal M}_4^{\rm p} + \nu  \left(  4 {\cal M}_4^{\rm t} \log\left(\tfrac{p_\infty}{2}\right)  +   {\cal M}_4^{\pi^2} + {\cal M}_4^{\rm rem}  \right) \right] 
		+ \int_{\bm \ell} \frac{\tilde{I}_{r,1}^4}{ Z_1  Z_2  Z_3 }
	+ \int_{\bm \ell} \frac{\tilde{I}_{r,1}^2 \tilde{I}_{r,2} }{ Z_1  Z_2 }
	+ \int_{\bm \ell} \frac{\tilde{I}_{r,1} \tilde{I}_{r,3}}{ Z_1} + \int_{\bm \ell} \frac{\tilde{I}_{r,2}^2}{ Z_1} \,,
	\nonumber \\[3pt]
	{\cal M}_4^{\rm p} &= 
%%%%% begin : M4pPaper
	- \frac{35 \left(1-18 \sigma ^2+33 \sigma ^4\right)}{8 \left(\sigma ^2-1\right)} 
%%%%% end : M4pPaper
	\,, \nonumber \\
	 {\cal M}_4^{\rm t} &= 
%%%%% begin : M4tPaper
	r_1 + r_2 \log \left(\tfrac{\sigma +1}{2}\right) +  r_3\frac{  {\rm arccosh} (\sigma) }{\sqrt{\sigma^2-1}}
%%%%% end : M4tPaper
	\,,
	\label{eq:amplitude} \\[3pt]
 {\cal M}_4^{\pi^2} &= 
%%%%% begin : M4Pi2Paper
	 r_4 \pi^2 + r_{5}\, \K \big(\tfrac{\sigma -1}{\sigma +1}\big) \E\big(\tfrac{\sigma -1}{\sigma +1}\big) 
                    + r_{6}\, \K^2\big(\tfrac{\sigma -1}{\sigma +1}\big)
                + r_{7}\, \E^2\big(\tfrac{\sigma -1}{\sigma +1}\big) \nonumber 
%%%%% end : M4Pi2Paper
 \,, \\[3pt]
                %%%
	{\cal M}_4^{\rm rem} &=  
%%%%% begin : M4remPaper
	r_8
	+ r_9 \log \left(\tfrac{\sigma +1}{2}\right) 
	+ r_{10}  \tfrac{ {\rm arccosh}(\sigma )   }{\sqrt{\sigma^2-1}}
	+ r_{11} \log (\sigma ) \nonumber 
	 + r_{12}\log^2 \left(\tfrac{\sigma +1}{2}\right)
	+r_{13}  \tfrac{\operatorname{arccosh}(\sigma) }{\sqrt{\sigma^2-1}} \log \left(\tfrac{\sigma +1}{2}\right)
	+ r_{14} \, \tfrac{ {\rm arccosh}^2(\sigma )}{\sigma^2-1}
	\\
	&\quad	+ r_{15}\,\text{Li}_2\left(\tfrac{1-\sigma }{2}\right)
	+  r_{16}\,\text{Li}_2\big(\tfrac{1-\sigma}{1+\sigma}\big)  \nonumber
    +  r_{17}\,\tfrac{1}{\sqrt{\sigma^2-1}}
	\left[ 
	\text{Li}_2\left(-\sqrt{\tfrac{\sigma-1}{\sigma +1}}\right)
	-\text{Li}_2\left(\sqrt{\tfrac{\sigma-1}{\sigma +1}}\right)\right] \nonumber   
%%%%% end : M4remPaper
	\,.
	\end{align}
	\vspace{-12pt}
\end{widetext}

This result is for the scattering of two scalars with incoming four-momenta $p_1$ and $p_2$ and masses $m_1$ and $m_2$.
Here $\bm{q}$ is the three-momentum transfer in the center of mass (COM) frame, $M = m_1 + m_2$ is the total mass, $\nu = m_1 m_2 /M^2$ is the symmetric mass ratio, $\sigma = p_1 \cdot p_2 / m_1 m_2$ in the mostly minus signature, and $p_{\infty}=\sqrt{\sigma^2-1}$.
For later convenience, we also introduce the three-momentum $\bm p_1= -\bm p_2= \bm p$ in the COM frame, individual energies $E_i=\sqrt{\bm p^2+m_i^2}$, the total energy $E=E_1+E_2$, and the symmetric energy ratio $\xi=E_1 E_2/E^2$.

The contribution in the probe limit is given by ${\cal M}_4^{\rm p}$, while the ${\cal O}(\nu)$ contribution, i.e. first order in the self-force expansion, consists of the logarithmic tail coefficient ${\cal M}_4^{\rm t}$, the potential $\pi^2$ contribution ${\cal M}_4^{\pi^2}$, and the remainder ${\cal M}_4^{\rm rem}$.

Note that the basis of transcendental functions has been slightly
rearranged with respect to Ref.~\cite{4PMPotential}, and that the
rational coefficients, given in Table~\ref{table:functions}, have
simplified upon combining the (ppp) and (prr) contributions. This
exposes additional structures; for example the coefficient of
$\operatorname{arccosh}^2\sigma$ is closely related to that of
$\operatorname{arccosh}\sigma$ in $\mathcal{M}_4^{\mathrm{t}}$,
suggesting that this expression exponentiates. Furthermore, the
coefficients of the weight two functions are expressible in terms of
the polynomial functions $g_1, g_2, g_3,$ and $g_4$. Note that the
function $g_3$ is related to the leading contribution of the electric
Weyl squared tidal operator to two-body scattering~\cite{DamourTidal}.

The remaining integrals in Eq.~(\ref{eq:amplitude}) are contributions
from iteration of lower order amplitudes. At ${\cal O}(G^4)$, these
arise only from the (ppp) region, and are thus the same as in our
previous work~\cite{4PMPotential}. Iteration terms in (prr), such as
from the last diagram in Fig.~\ref{fig:samplediagrams}, cancel in the
final amplitude. Similarly, terms of second order in the self-force
expansion as well as terms of transcendental weight three appear in
intermediate steps but cancel nontrivially in the final
amplitude. Note also that the PV prescription eliminates the imaginary
part of the amplitude, which is proportional to $\mathcal{M}_4^{\mathrm{t}}$.

%%%%%%%%%%%%%%%%%%%%%%%%%%%%%  TABLE %%%%%%%%%%%%%%%%%%
\begin{table}[tb]
	\setlength{\tabcolsep}{1pt} % Default value: 6pt
	\renewcommand{\arraystretch}{3}
	\begin{tabular}{|c|}
		\hline
		\scalebox{0.84}{\tabeq{10cm}{
				r_{1}  &= 
%%%%% begin : rPaper[1]
				\frac{1151 - 3336 \sigma + 3148 \sigma^2 - 912 \sigma^3 + 339 \sigma^4 
					- 552 \sigma^5 + 210 \sigma^6} {12(\sigma^2-1 ) }
%%%%% end : rPaper[1]
				\\
				r_{2}  &= 
%%%%% begin : rPaper[2]
				4g_1-2g_2-\frac{1}{2}g_3
%%%%% end : rPaper[2]
				\\
				r_3    &= 
%%%%% begin : rPaper[3]
				 \frac{\sigma\left(-3+2 \sigma^2\right)}{4(\sigma^2 - 1)}
				g_3
%%%%% end : rPaper[3]
				\\
				r_{4} &=
%%%%% begin : rPaper[4]
				-\frac{16}{3}g_1+\frac{4}{3} g_2 -\frac{1}{3}g_4
%%%%% end : rPaper[4]
				\\
				r_{5}   &=
%%%%% begin : rPaper[5]
				-\frac{1183 + 2929 \sigma + 2660 \sigma^2 + 1200 \sigma^3}{2 (\sigma^2-1 )} 
%%%%% end : rPaper[5]
				\\
				r_{6}   &=
%%%%% begin : rPaper[6]
				\frac{834+2095 \sigma +1200 \sigma^2}{2 (\sigma^2-1 )}
%%%%% end : rPaper[6]
				\\
				r_{7}   &= 
%%%%% begin : rPaper[7]
				\frac{7 \left(169 + 380 \sigma^2\right)}{4 (\sigma-1 )} 
%%%%% end : rPaper[7]
				\\
				r_8  &=
%%%%% begin : rPaper[8]
				\frac{1}{72(\sigma^2 -1)^2 \sigma ^7}\left(
				3600 \sigma^{16} + 4320 \sigma^{15} - 35360 \sigma^{14}+33249 \sigma^{13}\right.\\
				&\quad\left. \null + 27952 \sigma^{12} - 25145 \sigma^{11} - 15056 \sigma^{10} - 32177\sigma ^9
                                 + 64424 \sigma ^8\right.\\
				&\quad\left. \null -38135 \sigma ^7+13349 \sigma ^6-1471 \sigma^4 + 207 \sigma^2-45\right)
%%%%% end : rPaper[8]
				\\
				r_9  &= 
%%%%% begin : rPaper[9]
				-\frac{2 \left(75 \sigma^7 -  416 \sigma^5  - 612 \sigma^4 -  739 \sigma^3 
                                - 136 \sigma^2 -  2520 \sigma  -  152\right)} {3(\sigma^2-1)} 
%%%%% end : rPaper[9]
				\\
				r_{10}   &= 
%%%%% begin : rPaper[10]
				-\frac{\sigma (-3+2\sigma^2)}{(\sigma^2-1)} 2 r_1 + 28\sigma^2(-3+2\sigma^2)
%%%%% end : rPaper[10]
				\\
				r_{11}   &=
%%%%% begin : rPaper[11]
				\frac{4 \sigma}{3 (\sigma^2 - 1)} \left(-852 - 283 \sigma^2 - 140 \sigma^4 + 75 \sigma^6\right)
%%%%% end : rPaper[11]
				\\
				r_{12}   &= 
%%%%% begin : rPaper[12]
				6g_2+g_3-\frac{1}{2}g_4
%%%%% end : rPaper[12]
				\\
				r_{13}   &=
%%%%% begin : rPaper[13]
				-8\frac{\sigma (-3+2\sigma^2)}{\sigma^2-1} g_1
%%%%% end : rPaper[13]
				\\
				r_{14}   & =
%%%%% begin : rPaper[14]
				-\frac{\sigma^2(-3+2\sigma^2)^2}{4(\sigma^2-1)^2} g_3
%%%%% end : rPaper[14]
				\\
				r_{15}   &= 
%%%%% begin : rPaper[15]
				-16 g_1 - 4 g_2 - g_4
%%%%% end : rPaper[15]
				\\
				r_{16}   &= 
%%%%% begin : rPaper[16]
				-2g_4
%%%%% end : rPaper[16]
				\\
				r_{17}   &= 
%%%%% begin : rPaper[17]
				-\frac{\sigma (-3+2\sigma^2)}{\sigma^2-1} (8 g_1 - 4 g_2)
%%%%% end : rPaper[17]
				\\
				g_{1}&=
%%%%% begin : gPaper[1]
				2+15\sigma^2
%%%%% end : gPaper[1]
				\\
				g_{2}&=
%%%%% begin : gPaper[2]
				19\sigma+15\sigma^3
%%%%% end : gPaper[2]
				\\
				g_{3}&=
%%%%% begin : gPaper[3]
				11-30\sigma^2+35\sigma^4
%%%%% end : gPaper[3]
				\\
				g_{4}&=
%%%%% begin : gPaper[4]
				-20 -111 \sigma ^2 -30 \sigma ^4 + 25 \sigma^6
%%%%% end : gPaper[4]
		}}
\\
		\hline
	\end{tabular}
    \caption{Functions specifying the amplitude in \Eq{eq:amplitude}.
    }
	\label{table:functions}
\end{table}
%%%%%%%%%%%%%%%%%%%%%%%%%%%%% TABLE %%%%%%%%%%%%%%%%%%%%%%%%%%%%%%%%%%%%

As for the ${\cal O}(G^3)$ case, the result in \Eq{eq:amplitude} does not smoothly match onto
the massless case. Taking the high energy limit $|\bm p| \to \infty$, 
we find a leading power discontinuity in the classical part of the amplitude
\begin{align}
        {\cal M}_4^{\rm cons} = {560\pi^2(2- (2\ln2 + 1)^2) (G{\bm p}^2)^4  |\bm q| \over M \nu} + \dots,
\end{align}
where the ellipses denote terms subleading in large $|\bm p|$.  It would be
interesting to study whether this mass singularity cancels with dissipative effect
as it does at $\mathcal O(G^3)$~\cite{UltraRelativisticLimit}.

\sectionskip
\Section{Radial Action for Hyperbolic Orbits.} Given the amplitude-action relation~\cite{4PMPotential}, it is straightforward to derive the radial action for hyperbolic orbits 
from \Eq{eq:amplitude} via an inverse two-dimensional Fourier-transform. We find the radial action in the COM frame
\begin{align}
\label{eq:radial_4}
	I^{\rm hyp}_{r,4} &=
	%%%%% begin : Radial4
	-\frac{G^4 M^7 \nu^2\pi  {\bm p}^2}{8 E J^3} \\
	&\times \left[ {\cal M}_4^{\rm p} + \nu  \left(  4 {\cal M}_4^{\rm t} \log\left(\tfrac{p_\infty}{2}\right)  +   {\cal M}_4^{\pi^2} + {\cal M}_4^{\rm rem}  \right) \right]  \,, \nonumber 
\end{align}
which inherits the simple mass dependence from the amplitude~\cite{Bini:2020wpo,DamourSelfForce}. 
The radial action \eqref{eq:radial_4} in Mathematica text format is given in the ancillary file~\cite{AttachedFile}.
%. 
As an explicit benchmark, let us consider the nonlogarithmic (prr) contributions to the radial action defined in Ref.~\cite{BDGcons} as ${\cal M}_4^{\rm radgrav,f} \equiv
	14{\cal M}_4^{\rm t} + {\cal M}_4^{\rm rem,prr}$, with ${\cal M}_4^{\rm rem,prr}$ given in the Appendix. Expanding this through ninth PN order, we find
\begin{align}
	{\cal M}_4^{\rm radgrav,f}&=
	{12044 \over 75} p^2_\infty + {212077 \over 3675} p^4_\infty
	+{115917979 \over 793800} p^6_\infty \nonumber  \\
	&\quad - {9823091209 \over 76839840} p^8_\infty+{115240251793703 \over 1038874636800} p^{10}_\infty \nonumber \\
	&\quad -  {411188753665637 \over 4155498547200} p^{12}_\infty + \cdots \,, 
\end{align}
whose first three terms match the sixth PN order result in Eq.~(6.20) of Ref.~\cite{BDGcons}.

The scattering angle is given by $\chi_4=-\partial I_{r,4}^{\rm
  hyp}/\partial J$. After expanding in velocity, our result agrees
with the sixth PN order result in Eq.~(6.17) of
Ref.~\cite{BDGcons}. In particular, we find that terms of second order
in the self-force expansion are absent, consistent with
Ref.~\cite{BDGcons}.  On the other hand, the result of Ref.~\cite{Blumlein:2021txe} does
contain such a term (see Eq.~(69) therein). Aside from the
angle, another gauge-invariant observable is the time-delay for
hyperbolic orbits given by $\partial I_{r,4}^{\rm hyp}/\partial E$.

\sectionskip
\Section{Local Hamiltonian for Hyperbolic Orbits.} 
The presence of radiation leads to nonlocal-in-time contributions to
classical dynamics~\cite{TailEffect}. Following previous
analyses~\cite{2PM,3PMLong,4PMPotential}, a two-body Hamiltonian can
be derived from the radial action in Eq.~(\ref{eq:radial_4}).  
This Hamiltonian is effectively local and captures the nonlocal-in-time Hamiltonian evaluated on the relevant orbits.
For completeness, the effective local Hamiltonian valid for hyperbolic orbits is:
\eq{
H^{\rm hyp} = E_1+ E_2 + \sum_{n=1}^\infty {G^n  \over r^n} c_n(\bm p^2)\,,
}{eq:H}
where $r$ is the distance between the bodies in the COM frame. We have checked
that $H^{\rm hyp}$ reproduces the scattering angle. The lower-order
coefficients $c_1$, $c_2$, and $c_3$ can be found in Eq.~(10) of
Ref.~\cite{3PM}, while $c_4 = c_4^{\rm hyp}$ is
\begin{widetext}
\begin{align}
c_4^{\rm hyp} &=
%%%%% begin : c4Paper
 {M^7 \nu^2 \over 4\xi E^2} \left[ {\cal M}_4^{\rm p} + \nu {} \left(  4 {\cal M}_4^{\rm t} \log\left(\tfrac{p_\infty}{2}\right)  +   {\cal M}_4^{\pi^2} + {\cal M}_4^{\rm rem}  \right) \right] 
+ {\cal D}^3 \left[ {E^3 \xi^3 \over 3} c_1^4 \right] +{\cal D}^2 \left[ \left(\frac{E^3 \xi ^3}{\bm p^2}+\frac{E \xi {} (3 \xi-1 )}{2} \right) c_1^4 -2 E^2 \xi^2 c_1^2 c_2 \right] 
 \nonumber \\
&\quad 
+\left( {\cal D} + {1 \over \bm p^2} \right) \bigg[  E \xi {} (2c_1 c_3 + c_2^2) + \left( \frac{4\xi-1}{4E}+\frac{2 E^3 \xi ^3}{\bm p^4}+\frac{E \xi (3 \xi-1 )}{\bm p^2} \right)c_1^4 + \left((1-3 \xi )-\frac{4 E^2 \xi ^2}{\bm p^2}\right) c_1^2 c_2\bigg]
%%%%% end : c4Paper
\,,
\end{align}
\end{widetext}
where ${\cal D} = {d \over d \bm p^2}$ denotes differentiation with
respect to $\bm p^2$.  The final explicit result for $c_4^{\rm hyp}$
is included in the ancillary file~\cite{AttachedFile}. Again, the
label `hyp' emphasizes that $c_4^{\rm hyp}$ applies only for
hyperbolic orbits. Nonetheless, the potential contributions to ${\cal
  M}_4^{\pi^2}$ and the coefficient of logarithm, ${\cal M}_4^{\rm
  t}$, are local and should analytically continue between bound and
unbound orbits~\cite{Bini:2017wfr,Cho:2021arx}. 

\sectionskip
\Section{Conclusions.} 
We extended amplitudes methods to include the full conservative contributions
in classical two-body scattering at ${\cal O}(G^4)$. Compared to our
previous work on pure potential contributions~\cite{4PMPotential}, this
involves additional unitarity cuts and altered boundary conditions for
the differential equations that determine the master integrals. An
important next step will be to understand how to convert results for
scattering in the presence of conservative radiative effects to ones
for the bound-state problem~\cite{Bini:2017wfr,Cho:2021arx}.  It will
also be interesting to study the impact of our results on waveform
generation~\cite{BuonannoEnergetics}.

We expect that a setup similar to the one used at $\mathcal
O(G^3)$~\cite{HPRZ2} based on the formalism of Ref.~\cite{KMOC} can be
used to derive the dissipative radiation reaction contributions to the
scattering angle at $\mathcal O(G^4)$. Perhaps the most exciting
development is that we encountered no conceptual or technical
obstructions to pushing PM scattering calculations to increasingly
high orders and extracting from them the complete conservative
scattering dynamics of two-body systems.

\sectionskip
\Section{Acknowledgments.}                                                                 
We thank Johannes Bl\"umlein, Martin Bojowald, Alessandra Buonanno, Clifford
Cheung, Thibault Damour, Enrico Herrmann, Mohammed Khalil, David Kosower, Andr\'{e}s Luna,
 Andreas Maier, Philipp Maierhöfer, Aneesh Manohar, Peter Marquard, Rafael Porto, Ira Rothstein, Jan Steinhoff, Johann Usovitsch, and Justin Vines for helpful
discussions.  Z.B. is supported by the U.S. Department of Energy
(DOE) under award number DE-SC0009937.  J.P.-M.~is supported by the U.S.\
Department of Energy (DOE) under award number~DE-SC0011632.  R.R.~is supported
by the U.S.  Department of Energy (DOE) under award number~DE-SC00019066.
C.-H.S. is supported by the U.S.\ Department of Energy (DOE) under award
number~DE-SC0009919.   M.Z.'s work is supported by the U.K.\ Royal Society through Grant
URF{\textbackslash}R1{\textbackslash}20109.  We also are grateful to the
Mani L. Bhaumik Institute for Theoretical Physics for support.

\vskip .3 cm 

\appendix
\section{Amplitude in different regions.}

\vskip -.3 cm

%%%%%%%%%%%%%%%%%%%%%%%%%%%% TABLE %%%%%%%%%%%%%%%%%%%%%
\begin{table*}[t]
	\setlength{\tabcolsep}{1pt} % Default value: 6pt
	\renewcommand{\arraystretch}{3}
    \hfill
    \begin{minipage}{8cm}
    \centering
    \resizebox{8cm}{!}{
	\begin{tabular}{|c|}
		\hline
		\scalebox{0.84}{\tabeq{10cm}{
		r^{\rm ppp}_8  &= \vphantom{\frac{1}{2}}
%%%%% begin : rpppPaper[8]
        \frac{1}{144
   \bigl(\sigma^2-1 )^2 \sigma^7} (-45+207 \sigma^2-1471 \sigma^4+13349 \sigma^6 \\
& \quad - 37566 \sigma^7+104753 \sigma^8 - 12312 \sigma^9-102759 \sigma^{10}-105498 \sigma^{11} \\
   &\quad + 134745 \sigma^{12} + 83844 \sigma^{13} - 101979 \sigma^{14} + 13644 \sigma^{15} + 10800 \sigma^{16} \bigr)
%%%%% end : rpppPaper[8]
				\\
		r^{\rm ppp}_9  &= \vphantom{\frac{1}{2}}
%%%%% begin : rpppPaper[9]
\frac{1}{4 (\sigma^2 -1)} \bigl(
1759 - 4768 \sigma + 3407 \sigma^2 - 1316 \sigma^3 + 957 \sigma^4 \\
& \quad  - 672 \sigma^5 + 341 \sigma^6 + 100 \sigma^7 \bigr)
%%%%% end : rpppPaper[9]
			\\
        r^{\rm ppp}_{10}   &= \vphantom{\frac{1}{2}}
%%%%% begin : rpppPaper[10]
\frac{1}{24 (\sigma^2-1)^2}
\bigl(1237+7959 \sigma - 25183 \sigma^2 + 12915 \sigma^3 + 18102 \sigma^4 \\
& \quad - 12105 \sigma^5 - 9572 \sigma^6 + 2973 \sigma^7 + 5816 \sigma^8 - 2046 \sigma^9 \bigr)
%%%%% end : rpppPaper[10]
			\\
        r^{\rm ppp}_{11}   &= \vphantom{\frac{1}{2}}
%%%%% begin : rpppPaper[11]
 \frac{ 2 \sigma\left(-852 - 283 \sigma^2 - 140 \sigma^4 + 75 \sigma^6\right)}{3 (\sigma^2 - 1)}
%%%%% end : rpppPaper[11]
\\
        r^{\rm ppp}_{12}  &= \vphantom{\frac{1}{2}}
%%%%% begin : rpppPaper[12]
        4 g_1 - 7 g_2 - \frac{3}{4} g_4
%%%%% end : rpppPaper[12]
			\\
        r^{\rm ppp}_{13}  &= \vphantom{\frac{1}{2}}
%%%%% begin : rpppPaper[13]
        0 
%%%%% end : rpppPaper[13]
			\\
        r^{\rm ppp}_{14}   &= \vphantom{\frac{1}{2}}
%%%%% begin : rpppPaper[14]
        - \frac{\sigma^2\left(-3+2 \sigma^2\right)^2}{8(\sigma^2 - 1)^2}g_3 + 2(\sigma^2-1)g_2
%%%%% end : rpppPaper[14]
			\\
        r^{\rm ppp}_{15}  & = \vphantom{\frac{1}{2}}
%%%%% begin : rpppPaper[15]
		 24 g_1 - 14 g_2 + 2 g_3  - \frac{3}{2} g_4
%%%%% end : rpppPaper[15]
                \\
        r^{\rm ppp}_{16}  &= \vphantom{\frac{1}{2}}
%%%%% begin : rpppPaper[16]
			-g_4
%%%%% end : rpppPaper[16]
            \\
        r^{\rm ppp}_{17} &= \vphantom{\frac{1}{2}}
%%%%% begin : rpppPaper[17]
 -\frac{\sigma (-3+2\sigma^2)}{\sigma^2-1} (8 g_1 - 4 g_2)
%%%%% end : rpppPaper[17]
			\\
r^{\rm ppp}_{18} &= \vphantom{\frac{1}{2}}
%%%%% begin : rpppPaper[18]
  \frac{\sigma\left(-3+2 \sigma^2\right)}{2(\sigma^2 - 1)} \, g_3
%%%%% end : rpppPaper[18]
	}}
                \\
		\hline
	\end{tabular}
    }
	\caption{Functions specifying the amplitude in the ppp region.}
	\label{table:functionsppp}
    \end{minipage} \hfill
%\end{table*}
%
%\begin{table}[!hbt]
%	\setlength{\tabcolsep}{1pt} % Default value: 6pt
%	\renewcommand{\arraystretch}{3}
    \begin{minipage}{8cm}
    \centering
    \resizebox{8cm}{!}{
	\begin{tabular}{|c|}
		\hline
		\scalebox{0.84}{\tabeq{10cm}{
		r^{\rm prr}_8  &= \vphantom{\frac{1}{2}}
%%%%% begin : rprrPaper[8]
        \frac{1}{144 \sigma ^7 \left(\sigma ^2-1\right)^2} (-45+207 \sigma ^2-1471 \sigma ^4+13349 \sigma ^6        
\\
             &\quad
-38704 \sigma ^7+24095 \sigma ^8-52042 \sigma ^9+72647 \sigma ^{10}+55208 \sigma ^{11}
\\
             &\quad -78841 \sigma ^{12}-17346 \sigma ^{13}+31259 \sigma ^{14}-5004 \sigma ^{15}
 -3600 \sigma ^{16}
         )
%%%%% end : rprrPaper[8]
				\\
		r^{\rm prr}_9  &= \vphantom{\frac{1}{2}}
%%%%% begin : rprrPaper[9]
        \frac{1}{12 \left(\sigma ^2-1\right)} ( -4061 +34464 \sigma -9133 \sigma ^2
        +9860 \sigma ^3+2025 \sigma ^4 \\
             & \quad +5344 \sigma ^5-1023 \sigma ^6-900 \sigma ^7)
%%%%% end : rprrPaper[9]
			\\
        r^{\rm prr}_{10}   &= \vphantom{\frac{1}{2}}
%%%%% begin : rprrPaper[10]
\frac{1}{24 (\sigma ^2-1)^2} ( -1237 +5853 \sigma-16865 \sigma^2+15653 \sigma^3 + 
3018 \sigma^4  \\
& \quad -9011 \sigma^5 +5540 \sigma^6-3165 \sigma^7 -56 \sigma^8 + 366 \sigma^9
)
%%%%% end : rprrPaper[10]
			\\
        r^{\rm prr}_{11}   &= \vphantom{\frac{1}{2}}
%%%%% begin : rprrPaper[11]
        \frac{2 \sigma  \left(-852 -283 \sigma^2 -140 \sigma^4 + 75 \sigma^6\right)}{3( \sigma^2-1)}
%%%%% end : rprrPaper[11]
			\\
        r^{\rm prr}_{12}  &= \vphantom{\frac{1}{2}}
%%%%% begin : rprrPaper[12]
			    -4 g_1 + 13 g_2 + g_3 +\frac{1}{4} g_4
%%%%% end : rprrPaper[12]
			\\
        r^{\rm prr}_{13}  &=  \vphantom{\frac{1}{2}}
%%%%% begin : rprrPaper[13]
        - 8\frac{\sigma\left(-3+2 \sigma^2\right)}{(\sigma^2 - 1)} g_1 
%%%%% end : rprrPaper[13]
			\\
        r^{\rm prr}_{14}   &= \vphantom{\frac{1}{2}}
%%%%% begin : rprrPaper[14]
         - \frac{\sigma^2\left(-3+2 \sigma^2\right)^2}{8(\sigma^2 - 1)^2}g_3 - 2(\sigma^2-1)g_2
%%%%% end : rprrPaper[14]
			\\
        r^{\rm prr}_{15}  & =  \vphantom{\frac{1}{2}}
%%%%% begin : rprrPaper[15]
		 - 40 g_1 + 10 g_2 - 2 g_3  + \frac{1}{2} g_4
%%%%% end : rprrPaper[15]
                \\
        r^{\rm prr}_{16}  &= \vphantom{\frac{1}{2}}
%%%%% begin : rprrPaper[16]
			-g_4
%%%%% end : rprrPaper[16]
            \\
        r^{\rm prr}_{17} &= \vphantom{\frac{1}{2}}
%%%%% begin : rprrPaper[17]
0
%%%%% end : rprrPaper[17]
			\\
  r^{\rm prr}_{18} &= \vphantom{\frac{1}{2}}
%%%%% begin : rprrPaper[18]
 - \frac{\sigma\left(-3+2 \sigma^2\right)}{2(\sigma^2 - 1)} g_3
%%%%% end : rprrPaper[18]
	}}
                \\
		\hline
	\end{tabular}
    }
    \caption{Functions specifying the amplitude in the prr region.}
	\label{table:functionsprr}
\end{minipage}
%\end{table}
\end{table*}

In this appendix we present the scattering amplitudes in the (ppp) and (prr) regions separately. The contribution from the (ppp) region already appeared in Ref.~\cite{4PMPotential} and is given by
\begin{align}
&{\cal M}_4^{\mathrm{ppp}}(\bm q) = 
G^4 M^7 \nu^2 |\bm q|\pi^2 2^{2\epsilon} \left(\frac{ \bm q^2}{\tilde \mu^2}\right)^{-3\epsilon}  \nonumber
 \\
&\qquad\times
\left[{\cal M}_4^{\rm p} + \nu  \left( \frac{{\cal M}_4^{\rm t}}{\epsilon} + {\cal M}_4^{\pi^2} + {\cal M}_4^{\rm rem,ppp} \right)\right]   \\
    &
	\qquad + \int_{\bm \ell} \frac{\tilde{I}_{r,1}^4}{ Z_1  Z_2  Z_3 } + \int_{\bm \ell} \frac{\tilde{I}_{r,1}^2 \tilde{I}_{r,2} }{ Z_1  Z_2 }
	+ \int_{\bm \ell} \frac{\tilde{I}_{r,1} \tilde{I}_{r,3}}{ Z_1} + \int_{\bm \ell} \frac{\tilde{I}_{r,2}^2}{ Z_1}
	 \nonumber \,,
\end{align}
where ${\cal M}_4^{\rm p}$, ${\cal M}_4^{\pi^2}$, ${\cal M}_4^{\rm t}$ and the iteration integrals are as given in the main text in Eq.~\eqref{eq:amplitude}, and ${\cal M}_4^{\pi^2}+{\cal M}_4^{\rm rem,ppp}={\cal M}_4^{\rm f}$ defined in Eq.~(6) of Ref.~\cite{4PMPotential}.
The new result from the (prr) region is
\begin{align}
        {\cal M}_4^{\mathrm{prr}}(\bm q) &= 
        G^4 M^7 \nu^3 |\bm q|\pi^2 2^{6\epsilon} p_\infty^{-4\epsilon} \left(\frac{ \bm q^2}{\tilde \mu^2}\right)^{-3\epsilon} \nonumber\\ &\hspace{2cm}\times \left( -\frac{{\cal M}_4^{\rm t}}{\epsilon} +  {\cal M}_4^{\rm rem,prr}\right). 
  \label{eq:tailamplitude}
\end{align}

The remainder functions in both regions are given by
\begin{align}
	&{\cal M}_4^{\rm rem,x} =  
%%%%% begin : M4remxPaper
	  r^x_8
	+ r^x_9 \log \big(\tfrac{\sigma +1}{2}\big) 
    + r^x_{10}  \frac{ {\rm arccosh}(\sigma )   }{\sqrt{\sigma^2-1}}
    + r^x_{11} \log (\sigma ) 
	\nonumber \\
	&\quad  
    + r^x_{12} \log ^2\left(\tfrac{\sigma +1}{2}\right)
    + r^x_{13} \frac{\operatorname{arccosh}(\sigma) }{\sqrt{\sigma^2-1}} \log \left(\tfrac{\sigma +1}{2}\right) 
	\nonumber \\
	&\quad 
    + r^x_{14} \, \frac{ {\rm arccosh}^2(\sigma )}{\sigma^2-1} 
	+ r^x_{15} 
\text{Li}_2\left(\tfrac{1-\sigma }{2}\right)
	+ r^x_{16}
    \text{Li}_2\big(\tfrac{1-\sigma}{1+\sigma}\big)
	 \nonumber \\
	&\quad
    + r^x_{17}  \frac{1}{\sqrt{\sigma^2-1}}
	\left[ 
	\text{Li}_2\left(-\sqrt{\tfrac{\sigma-1}{\sigma +1}}\right)
	-\text{Li}_2\left(\sqrt{\tfrac{\sigma-1}{\sigma +1}}\right)\right]  \nonumber \\
    &\quad+ r^x_{18} \frac{1}{\sqrt{\sigma^2-1}} F(\sigma) {}
%%%%% end : M4remxPaper
	\,, \label{eq:M4remx}
\end{align}
where the relevant coefficients $r^x_i$ in each region, $x = {\rm ppp}$, ${\rm prr}$, are given in Tables \ref{table:functionsppp} and \ref{table:functionsprr} respectively in terms of the polynomials $g_i$ in Table ~\ref{table:functions}. The transcendental function $F(\sigma)$ in Eq.~\eqref{eq:M4remx} above is defined as
%\vskip -.5 cm
\begin{align}
         F(\sigma) &=
%%%%% begin : FPaper
\text{Li}_2\!\left(\!1\!-\!\sigma \!-\!\sqrt{\sigma ^2-1}\right)
        \! - \!  \text{Li}_2\!\left(\!1\!-\!\sigma \!+\!\sqrt{\sigma ^2-1}\right) \nonumber \\
        &+  3 \text{Li}_2\!\left(\!\sqrt{\tfrac{\sigma-1}{\sigma +1}}\right)
        \! - \!  3 \text{Li}_2\!\left(\!-\sqrt{\tfrac{\sigma-1}{\sigma +1}}\right)  \nonumber \\
        &+  2 \log \left(\tfrac{\sigma +1}{2}\right) {\rm arccosh} (\sigma)
%%%%% end : FPaper
\,,
\label{eq:Ffunction}
\end{align}
and its coefficients cancel when both regions are combined.

%%%%%%%%%%%%%%%%%%%%%%%%%%

\end{document}